\documentclass{amsart} 

\usepackage{amsmath}
\usepackage{hyperref}
\usepackage{gensymb} 

\usepackage{algorithmic}

\usepackage{natbib}

\usepackage{graphicx}
\usepackage{longtable}
\usepackage{multirow}
\usepackage{eurosym}
\usepackage{xcolor,colortbl}
\usepackage{amsmath}
\usepackage{float}
\usepackage[caption = false]{subfig}
\definecolor{Gray}{gray}{0.85}
\newcolumntype{a}{>{\columncolor{Gray}}p}
\begin{document}

		\title{Nanosatellite aerobrake maneuvering device}

		%% Group authors per affiliation:
		\author{Valeriia G. Melnikova}
		\email{vg-melnikova@yandex.ru}
		
		\author{Alexander A. Borovikov}
		\email{borovic68@mail.ru}
		
		\author{Maksim I. Koretskii}
		\email{hardsofa@gmail.com}
		
		\author{Yuliya L. Smirnova}
		\email{yuliya.sm7@gmail.com}
		
		\author{Ekaterina D. Timakova}
		\email{ti-kate7@yandex.ru}
		
		\author{Zhaokai Yu}
		\email{yuzhaokai933@gmail.com}
		
		\author{Arseniy O. Kuznetsov}
		\email{arseniy.kuznetsov.ejje@gmail.com}
		
		\author{Kirill A. Frolov}
		\email{dinamik1994@yandex.ru}
		
		\author{Stepan M. Tenenbaum}
		\email{ivankovo@list.ru}
		
		\author{Dmitriy A. Rachkin}
		\email{rachkin@bmstu.ru}
		
		\author{Oleg S. Kotsur}
		\email{oskotsur@gmail.com}
		
		\author{Nikolay A. Nerovny}
		\email{nick.nerovny@bmstu.ru}
		
		\author{Vera I. Mayorova}
		\email{victoria.mayorova@gmail.com}
		
		\author{Anton S. Grigorjev}
		\email{baldy.ash@yandex.ru}
		
		\author{Nikita V. Goncharov}
		\email{layreat@yandex.ru}

		\address{Bauman Moscow State Technical University, 5 stroenie 1 2-ya Baumanstaya st., Moscow, 105005, Russia}
		
		\begin{abstract}
			In this paper, we present the project of the heliogyro solar sail unit for deployment of CubeSat constellation and satellite deorbiting. The ballistic calculations show that constellation deployment period can vary from 0.18 years for 450km initial orbit and 2 CubeSats up to 1.4 years for 650km initial orbit and 8 CubeSats. We also describe the structural and electrical design of the unit and consider aspects of its integration into a standard CubeSat frame.
			
		\end{abstract}

	\maketitle
	
	\allowdisplaybreaks

\markleft{V. MELNIKOVA ET AL.}
	
\section{Introduction}
The rising century of small satellites creates more challenging tasks for them. One of the development trends of small satellites is the  creation and deployment of CubeSat~\cite{heidt_cubesat:_2000} constellations to implement some distributed activities: remote sensing, including Earth observation and climate monitoring~\cite{esper_nasa-gsfc_2000,sandau_small_2010}, Sun observation, and space weather~\cite{robinson_small_2008}, communications~\cite{bedon_preliminary_2010}, ionospheric research~\cite{sandau_small_2010-1}, and other activities.
Several projects have already been announced, some of them have started space operations: QB50\footnote{\url{https://www.qb50.eu/}}~\cite{gill_formation_2013}, Planet Labs\footnote{\url{https://www.planet.com/}}~\cite{marshall_planet_2013,boshuizen_results_2014}, Spire\footnote{\url{https://spire.com/}}~\cite{borowitz_is_2015}, and others are coming~\cite{barnhart_very-small-satellite_2007,barnhart_low-cost_2009,bandyopadhyay_review_2015-1,crisp_launch_2015}.

In the case of deploying of satellites at the same time from the same launcher, it needs a propulsion system to deploy a constellation. Another way is to launch CubeSats in different time during several months from the same launching satellite (e.g., as Planet Labs do on ISS). The disadvantage of launch from the ISS is the growing period of constellation formation and a considerable atmospheric drag which reduce the operational time of the satellites. The advantage of this option is a possible reduction of launch failures, so there is a trade-off between potential risk value and an operational lifetime.

During the AeroCube4 mission\footnote{\url{https://directory.eoportal.org/web/eoportal/satellite-missions/a/aerocube-4}}~\cite{gangestad_operations_2013}, the satellite successfully implemented several orbital maneuvers using atmospheric drag forces, but the problem of constellation deployment was not solved.
There are several studies with the similar method for constellation deployment using atmospheric drag forces~\cite{li_optimal_2014,foster_orbit_2015}.
It is also possible to utilize this technology for spacecraft de-orbiting -- as it is already implemented now\footnote{\url{http://www.sstl.co.uk/Blog/September-2013/The-De-Orbit-Sail-that-will-bring-TechDemoSat-1-ou}}~\cite{alhorn_nanosail-d_2011,ridenoure2015lightsail}.

Deployment of several CubeSats on the orbit using drag forces can be accomplished within several months. The duration of deployment depends on orbit altitude and drag area. We should notice that any active CubeSat propulsion system will reduce the deployment time to the days and even hours~\cite{schmuland_mission_2012,carpenter_cubesat_2013}, and this is the significant disadvantage of any design of the satellite equipment which utilizes atmospheric drag force for passive formation deployment.

In this paper, we proposed a method for formation of satellite constellation using solar sail technology.
A solar sail uses light pressure to create thrust without consumption of propellant.
There were several successful experiments with solar sails on CubeSats, namely Nanosail-D2~\cite{alhorn_nanosail-d_2011,johnson_nanosail-d:_2011} and Lightsail-1~\cite{ridenoure2015lightsail}.
For the low Earth orbit, light radiation pressure is lower than atmospheric drag~\cite{mcinnes_solar_2004}.
However, the similar effect of air drag on the sail can be used to deploy a satellite constellation at altitudes lower than 550 km.

One of the types of solar sails is a Heliogyro -- rotary solar sail~\cite{macneal_1967}.
Recent progress in the field of Heliogyro solar sails~\cite{wilkie_recent_2014,guerrant_tactics,heiligers_heliogyro_2015,bmstu_iac_2015,NerovnyBMSTUSailSpaceExperiment2017} allows us to utilize this technology for practical space applications.

\section{The constellation deployment algorithm}

The operating principle is a changing of the atmospheric drag area. By utilizing of Heliogyro-like solar sail, it is possible to roll-out and roll-in the sail blades onto the bobbins.

Let us consider the problem of separation of $N$ satellites from the common initial position on the orbit. For every satellite, we know the geometrical parameters of its surface and its weight. All spacecraft should be dispersed evenly throughout the orbit. The final orbit has to be close to the starting orbit.
At the initial time $t=0$, all satellites remain approximately at the same point of the orbit. The satellite number $1$ deploys its solar sail, which results in changes of parameters of its orbit.
After this, the phase mismatch $\Delta \phi_1$ between satellite $1$ and the other satellites will start to increase.
Next, after a particular time, satellite $1$ folds the sail, and the next satellite $2$ deploys its sail. Then the satellite $2$ appears on the orbit which minimizes the relative movement of both spacecraft, keeping a constant angle $\Delta \phi_0$ between the satellites, and the sail on the satellite $2$ folds up.
This procedure continues up to the satellite $N-1$.
The last satellite should also deploy its sail for a particular time $\Delta T_N$ to achieve the final orbit in which phase mismatch with all other satellites will remain approximately constant.

This constellation deployment procedure can be described as the following algorithm.

\begin{algorithmic}
	\REQUIRE $N > 0 \vee \{\Delta \phi_k = 0 | k=1,\dots,N-1\} \vee \Delta \phi_0>0$
	\ENSURE $\Delta \phi_k \cong \Delta \phi_0$
	\STATE $t \leftarrow 0$
	\FOR{$k \in \{1,\dots,N\}$}
	\STATE $t_k \leftarrow t$
	\STATE Deploy sail for satellite $k$
	\IF{$k\neq N$}
	\STATE Wait $\Delta T_k$ until $\Delta \phi_k \cong \Delta \phi_0$
	\ELSE[$k=N$]
	\STATE Wait $\Delta T_N$
	\ENDIF
	\STATE $T_k \leftarrow t + \Delta T_k$
	\STATE $t \leftarrow T_k$
	\STATE Fold sail for satellite $k$%
	\ENDFOR
\end{algorithmic}

The particular values of $\Delta T_k$ may be acquired by a multidimensional optimization.
The optimization condition is the fastest time of formation deployment $T_N$.
The optimization parameters are the time of start of deployment for each sail ($t_k$) and the period for which the particular solar sail will be opened ($\Delta T_k$).

\section{Modeling of constellation deployment}

Let us consider the Earth-centered inertial equatorial coordinate system for the formulation of differential equations of the movement of the center of mass of spacecraft.

These are the general assumptions we used for the differential equations of motion:
\begin{itemize}
	\item all external forces on the spacecraft acts to its center of masses; there is no external torque from gravitational and atmospheric forces;
	\item we neglect gravitation of the Moon, Sun, and other planets;
	\item the atmosphere of the Earth does not rotate;
	\item the solar sail maintained its initial orientation during the calculation time, which is typical for a rotary solar sail, stabilized by rotation.
\end{itemize}

The differential equations of motion of the spacecraft have the following simple form:
\begin{equation}
\frac{d^2\mathbf{X}}{dt^2} = \frac{1}{m}(\mathbf{F}_g+\mathbf{F}_a+\mathbf{F}_s),\label{eq:newton}
\end{equation}
Where: $\mathbf{X}$ -- vector of coordinates of the satellite; $m$ -- the weight of satellite; $\mathbf{F}_g$ - vector of the gravitation force of the Earth, in which model we take into account only a second zonal harmonic; $\mathbf{F}_a$ - vector of an atmospheric drag force; $\mathbf{F}_s$ - vector of a solar radiation pressure force.

The atmospheric drag pressure calculated by existing mathematical models (Eq.~(\ref{eq:newton})) is very uncertain and depends on many factors defined by mission profile (attitude, solar activity index, Earth penumbra effects, sail orientation, and its shape).

We used these conservative assumptions for the model of the atmosphere: 
\begin{itemize}
	\item solar activity index $F_{10.7} = 100$ sfu for the 2016\dots2022 as NOAA prognoses\footnote{\url{http://services.swpc.noaa.gov/ images/solar-cycle-10-cm-radio-flux.gif}};
	\item flow is normal to sail surface;
	\item atmospheric drag coefficient $C_x = 2.2$.
\end{itemize}

\begin{figure} \centering\includegraphics[width=80mm]{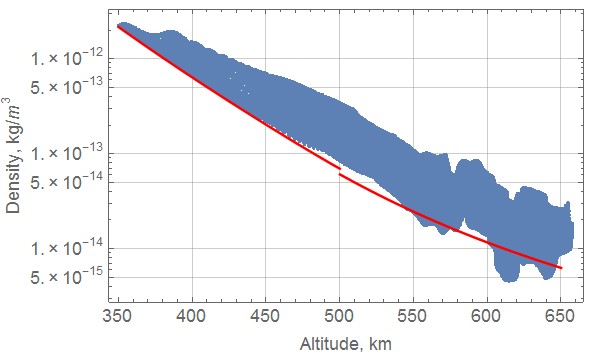}
	\caption{Comparison of GOST R 25645.166-2004 model for density of upper atmosphere (red line, for shadowed part of orbit) with MSIS-E-90 atmosphere model (blue zone)} \label{fig:gost}
\end{figure}

We used the GOST R 25645.166-2004 atmospheric model~\cite{golikov_theonanumerical-analytical_2012}. We neglected the effects of Earth penumbra.
The comparison between GOST atmosphere and MISI-E-90 model\footnote{\url{http://omniweb.gsfc.nasa.gov/vitmo/msis_vitmo.html}}~\cite{bilitza2002msis} is presented in the Fig.~\ref{fig:gost}.
The comparison was made in GMAT 2015a software package\footnote{\url{http://gmatcentral.org/display/GW/2015/11/02/Announcing+GMAT+R2015a}}.

The solar radiation pressure was assumed to be $0.9\cdot10^{-6}$ Pa, and the sail is flat and entirely specularly reflective.
The solar radiation flux was also assumed to be perpendicular to the sail surface (cannonball model). Currently, we are investigating the improved model of solar radiation pressure based on the modification of the Generalized Sail Model~\cite{rios-reyes_generalized_2005,nerovny_representation_2017}.

\paragraph{Initial orbit}
inclination $51.6\degree$ (ISS orbit inclination), longitude of the ascending node $0\degree$, the argument of periapsis $0\degree$, eccentricity $0.0001$, true anomaly $0\degree$, a various initial height of the orbit.

\paragraph{Satellite parameters}
The weight of each nanosatellite is always equal to 3kg.
This assumption corresponds to the average weight of the 3U CubeSat satellite.
The initial height of the orbit changed from 450 km to 650 km with step 50 km, the drag area of solar sail changed from 0.135 $m^2$ to 1.01 $m^2$, and the number of satellites was assumed to be 2, 4, or 8.    

\paragraph{Optimization}
Starting from the maximum possible time span, which is equal to 1000 seconds, we looked for the value of time of deployment of next solar sail $t_i$ using golden section method. After this, the software calculated values of $T_i$ also using golden section method.
The optimization stopped when the deficiency of the calculated and predicted phase separation angles became less than proposed accuracy level.

\begin{figure} \centering\includegraphics[width=80mm]{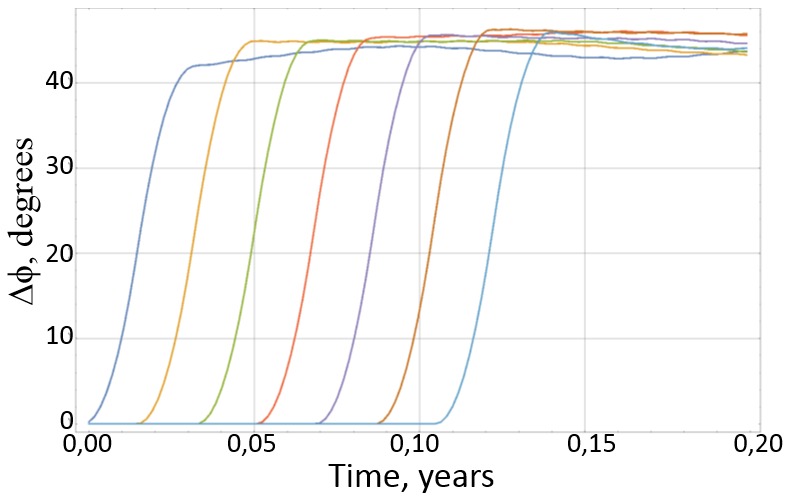}
	\caption{Evolution of the phase angles between 8 spacecrafts with the initial height of orbit 450 km and drag area 1.01 $m^2$} \label{fig:evolution}
\end{figure}

\begin{figure} \centering\includegraphics[width=80mm]{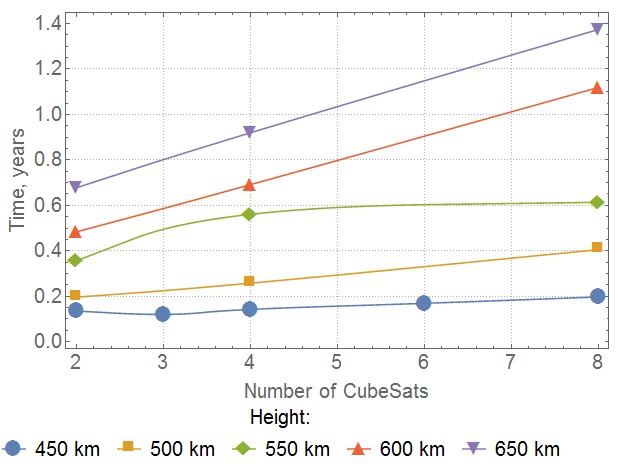}
	\caption{Influence of height for the period of deploying (drag area -- 0,51 $m^2$).} \label{fig:influence}
\end{figure}

Fig.~\ref{fig:evolution} shows that the phase angle between spacecraft is stabilized and becomes constant after the separation. Thus the constellation is formed, confirming the feasibility of the concept. The divergences at the end of simulation time in the figure depend on the accuracy of the calculation, and it is possible to reduce them with the increase of the accuracy of the simulation.
As shown in the figure, with the increase of an altitude of the initial orbit the average density of the atmosphere decreases, and the time of constellation formation increases.

\begin{figure} \centering\includegraphics[width=80mm]{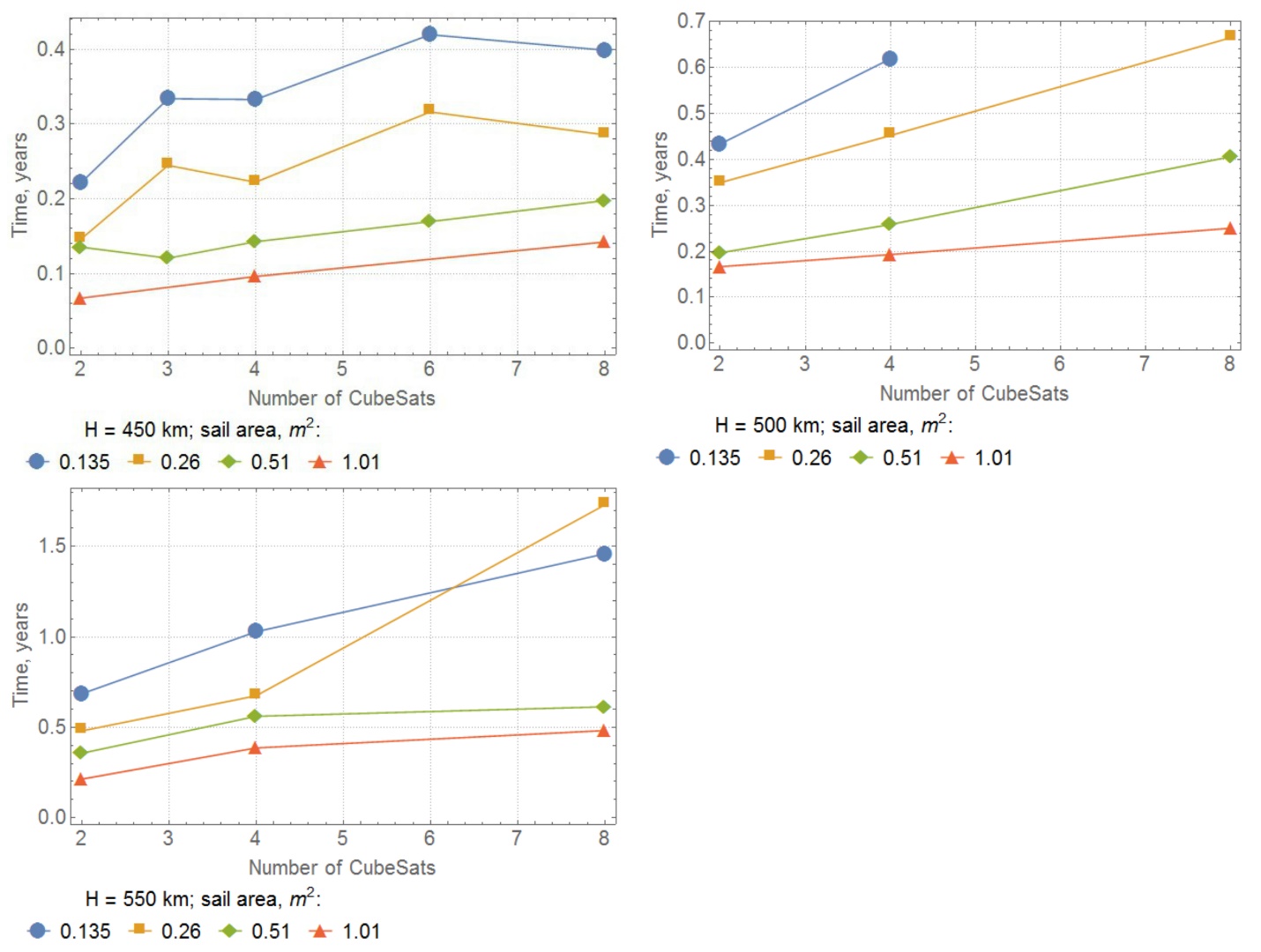}
	\caption{Influence of the drag area of the sail for the time of constellation deployment. Respectively, then the more the drag area of the sail, constellation occurs the quicker.} \label{fig:influence_2}
\end{figure}

All carried calculations showed that cumulative decrease of an altitude of the orbit does not exceed 15 km. The decline of satellite lifetime is almost equal to the duration of constellation deployment for low attitude orbit (450 km) and become negligible for altitudes higher approximately 550 km (Fig.~\ref{fig:influence} and~\ref{fig:influence_2}).

\section{Solar sail deployment}

The nanosatellite should spin around some axis perpendicular to the sail surface to provide stiffness to the sail.    
In this case, sail stiffness has to be enough to deploy sail and to withstand atmospheric drag pressure and solar pressure.    
According to the results from BMSTU-Sail deployment simulations~\cite{NerovnyBMSTUSailSpaceExperiment2017}, there is a relationship between initial spinning velocity and the required sail deployment velocity. This relationship can be obtained by the numerical simulations of the sail deployment process considering it as a tether~\cite{bmstu_iac_2012,tenenbaum_2014}.

The duration of sail deployment may differ from hundreds of seconds to the thousands of seconds or some hours, but in all cases, the deployment time is negligible compared with nanosatellite lifetime (months or years). 

\begin{figure} \centering\includegraphics[width=80mm]{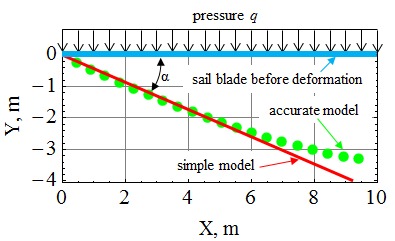}
	\caption{Sail blade deformation under atmospheric drag and solar pressure.} \label{fig:sail}
\end{figure}

The solar radiation pressure and drag pressure affect the value of spinning velocity because they deform the shape of the solar sail.
We introduced the angle $\alpha$ between sail tip point before deformation, central body and sail tip position after deformation as a numerical measure of such deformation (Fig.~\ref{fig:sail}). 
For normal sail operation, $\alpha$ has to be as low as possible to prevent the decrease of the effective area of the solar sail. 
For this paper, we used condition $\alpha\leq10\degree$  which provides effective sail area about 95\% from its overall area.
Simulations of sail deformations indicated that at $\alpha<20\degree$, sail deformed shape is close to the straight line (Fig.~\ref{fig:sail}). In that case, one can derive the equation for $\alpha$ from equilibrium equation of torques from combined atmospheric and solar pressure, and centrifugal forces:
\begin{equation}
\alpha = \arctan\frac{q}{\left( \frac{2}{3} m_{sail} + 2m_{tip} \right) \Omega^2 },\label{eq:alpha}
\end{equation}
Where: $\alpha$ -- deformation angle, $radians$ (Fig.~\ref{fig:sail}); $q$ -- combined atmospheric drag and solar pressure on one meter of blade length, $N/m$; $m_{sail}$ -- weight of one sail blade, $kg$; $m_{tip}$ -- sail blade tip mass; $\Omega$ -- spinning velocity of satellite with deployed sail, $radians/s$.
Comparing the exact solution for sail deformation, the error extimation for Eq.~(\ref{eq:alpha}) is about 1$\degree$ that is negligible.

\begin{figure} \centering\includegraphics[width=80mm]{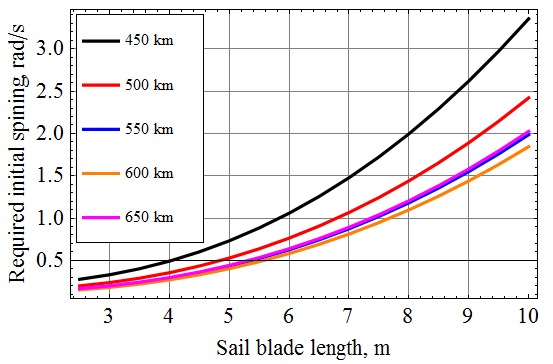}
	\caption{Required nanosatellite initial spinning velocity as the function of sail blade length and orbit altitude (from Eq.~(\ref{eq:omega})).} \label{fig:velocity}
\end{figure}

We can find the relationship between spinning velocity of deployed sail from Eq.~(\ref{eq:alpha}) and the required initial spinning velocity before deployment using the principle of conservation of angular momentum:
\begin{equation}
\Omega_0 = \Omega \left( 1 + \frac{Y_{sail}}{Y_{CubeSat}} \right),\label{eq:omega}
\end{equation}
Where: $\Omega_0$ -- required initial spinning velocity; $Y_{sail}$ -- inertia moment of deployed sail about spinning axis; $Y_{CubeSat}$ -- inertia moment of CubeSat about spinning axis before deploying. 
Fig. \ref{fig:velocity} represents the results of calculations of initial spinning velocity for different flight altitudes.

\begin{figure} \centering\includegraphics[width=80mm]{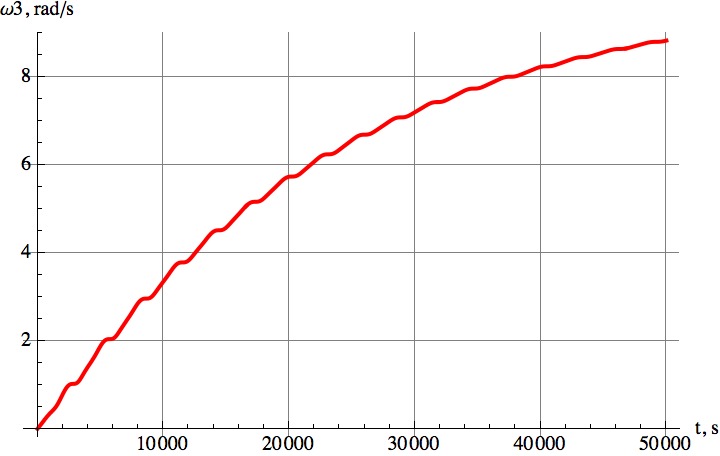}
	\caption{Angular velocity of CubeSat.} \label{fig:angular}
\end{figure}

We conducted the calculations of the spin-up time considering the standard magnetic attitude control system\footnote{\url{http://www.cubesatshop.com/product/cubetorquer-and-cubecoil/}} and the known control laws~\cite{ilyin_et_al_2005}, Fig.~\ref{fig:angular}.
For the range of the possible sail surface area and altitudes of the orbit, the required initial angular velocity varies from 0.2 to 4.0 rad/s.
The commercially available CubeSat magnetic torques can provide this spin-up. Taking into account the various constraints on the CubeSat, such as power and link budget, the spin-up duration can last up to the several days at maximum.
This period is much lower comparing with nanosatellite lifetime and the length of constellation deployment time by utilizing of proposed in this paper technology.

\section{Solar sail unit}

\subsection{Operation}

The ground station guides the deployment of a constellation. 
The guidance consists of commands for opening and closing of a solar sail for every nanosatellite at the different time.
We choose the following separation of operational functions between nanosatellite bus and a solar sail unit:
\begin{itemize}
	\item nanosatellite provides power supply;
	\item nanosatellite provides commands from ground station for deploying/folding of the sail using the standard digital interface;
	\item nanosatellite provides spinning with the necessary angular velocity and necessary attitude, e.g., by using of magnetic coils;
	\item the sailing unit deploys or folds the sail;
	\item the sailing unit provides telemetry to the nanosatellite which is relaying it to the ground station.
\end{itemize}

Solar sail unit is operating in the hibernation mode at the time when deploying/folding of the sail does not need.

\subsection{General design}

\paragraph{Design requirements}
The main requirements for the unit layout are the following:
\begin{itemize}
	\item Dimensions of all of the PCBs with electronic components has to be 96 x 90 mm;
	\item Height of the unit should be minimized to provide additional space for a payload, but enough to be connected with the other parts of the CubeSat using electrical interfaces;
	\item Width of the sail should be as large as possible to minimize the length of the sail;
	\item Bobbin's structure should be designed in two versions: one made from aluminum with a milling machine, and the other should be polyamide;
	\item It has to be possible to make several structural elements with a 3D printer;
	\item All of the designed components has to be durable and reliable enough to withstand the loads during the launch and subsequent flight.
\end{itemize}

\begin{figure} \centering\includegraphics[width=80mm]{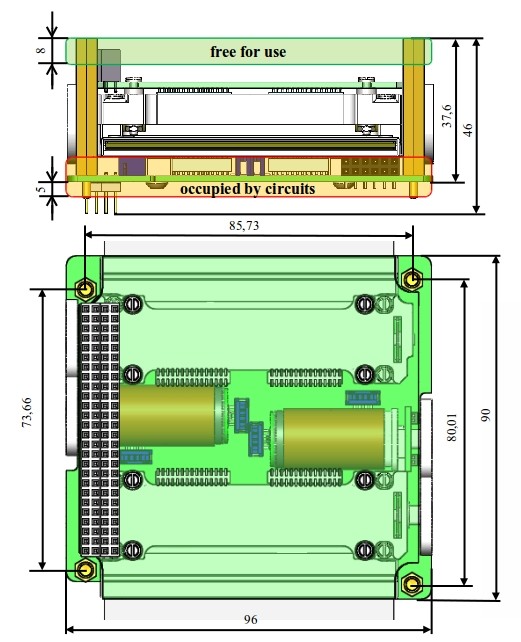}
	\caption{Solar sail unit dimensions.} \label{fig:solar}
\end{figure}

\begin{figure} \centering\includegraphics[width=80mm]{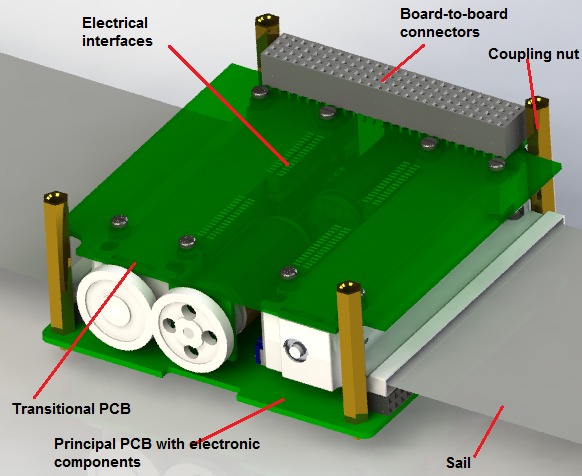}
	\caption{Unit layout.} \label{fig:unit}
\end{figure}    

\paragraph{Unit}

\begin{figure} \centering\includegraphics[width=80mm]{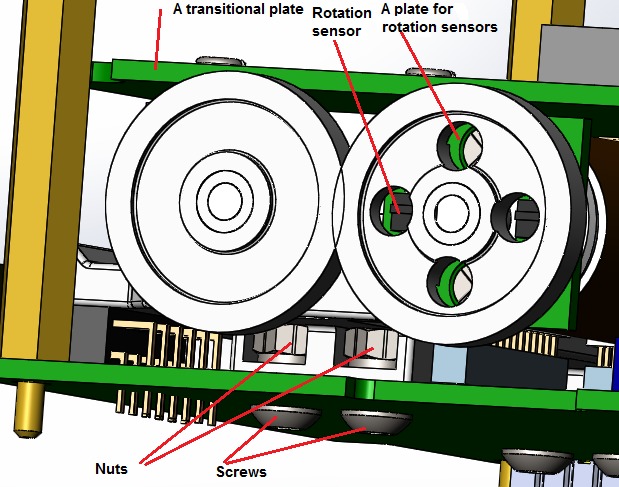}
	\caption{Unit structure.} \label{fig:unit_2}
\end{figure}

Fig.~\ref{fig:solar} shows the dimensions of the unit.
The layout of the unit constrained by the sail width which depends on the distance between the coupling nuts (Fig.~\ref{fig:unit}).
The location of the bobbins is symmetrical relative to the spinning axis of the satellite; they should be close to the center of the unit mass.
The fixture of the bobbins to the main plate is eight screws locked with nuts and washers. The materials of all the fittings are non-magnetic stainless steel. The PCB has micro motion sensors mounted on a gear axis behind the motor wheel, and there are specific holes in the wheel which help to register the spinning velocity of the bobbins (Fig.~\ref{fig:unit_2}).

\paragraph{Bobbin}

\begin{figure} \centering\includegraphics[width=80mm]{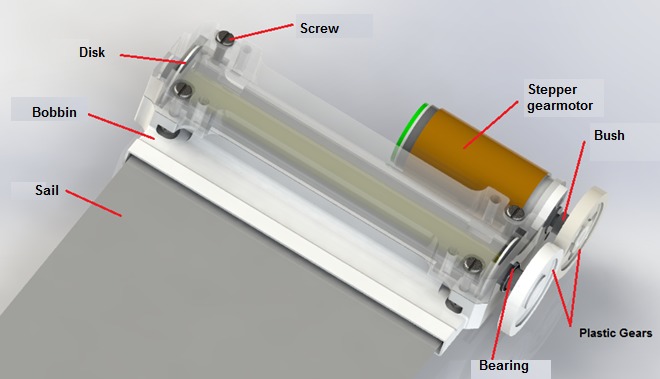}
	\caption{Bobbin with a polyamide body.} \label{fig:bobbin}
\end{figure}

A bobbin body design has two different variants: one for manufacturing on a 3D printer, and the other for manufacturing on a milling machine.    
At Fig.~\ref{fig:bobbin} one can see a bobbin with a polyamide body consisting of two parts and the middle parting plane in between. The top of the body is fixed to the bottom with four screws. The axis and the disks preventing the sail from contacting the bearings while deploying or folding are printed as one.  The plastic gears are mounted on printed polyamide bushes. We decided not to make a thread from polyamide because it is flexible and can damage the sail film.

\begin{figure} \centering\includegraphics[width=80mm]{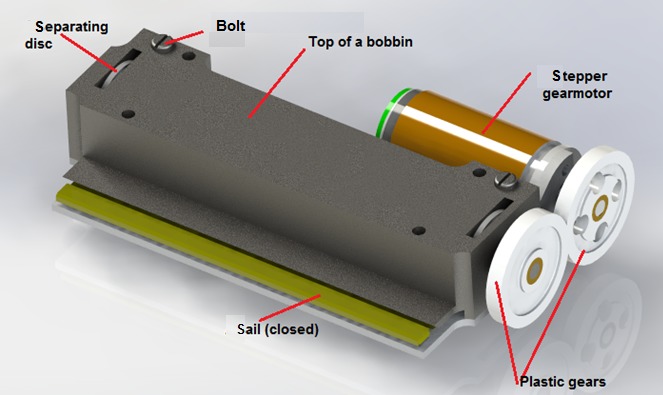}
	\caption{Unit with a duralumin bobbin.} \label{fig:bobbin_3}
\end{figure}

The other variant of the bobbin is the variant in which the bobbin body is made of duralumin (Fig.~\ref{fig:bobbin_3}).
The body parts are fixed together with three bolts.

Comparison of these two variants showed that the second one seems to be more reliable and durable. However, the 3D printing technology allows making things appropriate for use in space. Firstly, the polyamide structure is much lighter than an aluminum one, and there can be an extra payload on a board of the CubeSat.  Secondly, it is much easier and cheaper in production comparing traditional milling technology. Thirdly, the polyamide does not dissolve during the flight, and finally, it proved to be durable and hard enough to sustain the loads during the launch and detachment. There is only one problem with polyamide -- it should be metalized to prevent charging and ESD effects in the space environment. This problem can be solved by using specific paintings on the outer surfaces of bobbin body.

\subsection{Integration}

\begin{figure} \centering\includegraphics[width=80mm]{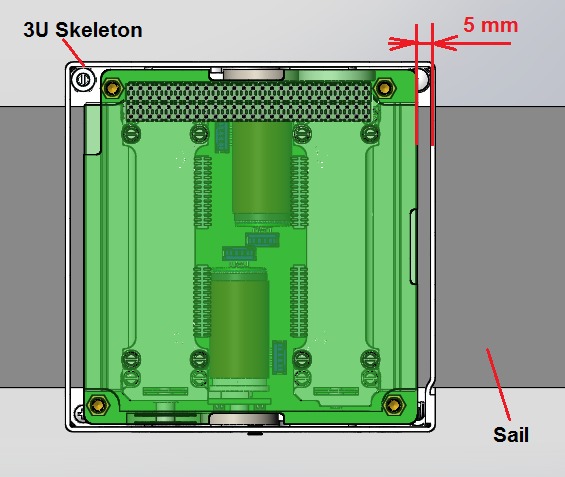}
	\caption{Solar sail unit integration into 3U CubeSat. 
		Gap between the unit and the CubeSat skeleton.} \label{fig:solar_2}
\end{figure}

There are several requirements to the sailing unit for its integration into the CubeSat:
\begin{itemize}
	\item The outer dimensions of the unit are to be no larger than 90mm x 96mm x 37mm;
	\item It should provide as much as possible of free space to the CubeSat;
	\item It should afford the possibility of its placement in the middle of CubeSat stack as well as on the periphery including by-passing of electrical interfaces;
	\item There must be cutouts for the sail in the CubeSat skeleton which provide the 1mm clearance between the sail and the skeleton;
	\item It should include the special adapter which closes the gap between the sail and the CubeSat skeleton (Fig.~\ref{fig:solar_2}). It is proposed to attach this adapter to the solar panels with glue after the unit integration.
\end{itemize}

\begin{figure} \centering\includegraphics[width=80mm]{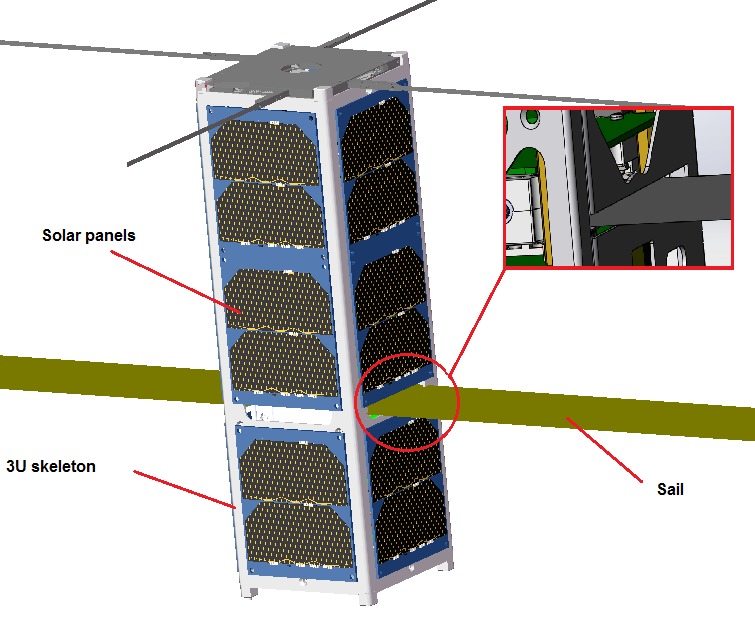}
	\caption{The integration of solar sail unit into CubeSat.} \label{fig:interation}
\end{figure}

The example of solar sail unit integration to 3U CubeSat is shown in Fig.~\ref{fig:interation}.
For this paper, we used the standard CubeSat components from CubeSatShop\footnote{\url{http://www.cubesatshop.com/faq/}}, however, it is proposed that this unit can be adapted to any other CubeSat-class satellite.

Electrical interfaces in different CubeSats have similar principles, although practical implementation differs. At the current stage, we selected the ISIS interface, which includes two board-to-board connectors.

To provide the solar sail unit functionality, we used I2C for main and reserve information buses, and one power feeder (3.0 to 5.0 V).

\section{Control system}

The solar sailing unit is equipped with its avionics which uses the CubeSat's electricity supply, data, and command link with the ground station through CubeSat’s transceiver.
The solar sail unit acts as a slave device on the I2C bus.
The avionics has to be reliable; it should be tolerant to one permanent fault. It was proposed to utilize the  Russian components of industrial class primarily. It should work on the broad allowable range of input/output voltage levels. 
Avionics consist from DC-DC converter, microcontrollers,  MEMS sensors, FLASH memory, input/output ports, etc.

The DC-DC converter provides stabilized power buses (3,3 V \& 5,0 V) for unit operation in the broad range of input voltage. There is input filter (capacitance) and protection (TVS diode) at the input power line from CubeSat.
It consists of several DC-DC devices based on KP1156EY chip (STC SIT) and is constructed by the cascade principle. The first step-up works in the voltage range from 3 to 5 V and transform it to the stable 5 V. This line provides power for some failsafe subsystems, for the second stage of DC-DC and other components. The second stage of the DC-DC converter is a step-down converter. It provides the stable voltage of 3.3 V for main integrated circuits. This scheme ensures the operation of a unit at input voltage range from 3 to 5 V. Each stage of the DC-DC has hot redundancy, and there is protection from current overload.

Two I2C interfaces provide data and command handling. There is a bidirectional optical isolation on the entrance of the interface for more reliability. The I2C bus usually reserved on a board of CubeSat that is why these parts have no redundancy. 
Internal electronic equipment operates on another two separate I2C buses.

We selected the industrial-class version of Russian microcontroller K1986BE92QI by Milandr based on the ARM Cortex-M3 architecture as a core of the avionics.
Failure of the microcontroller is critical for all the system, that is why we use the scheme of cold redundancy. This scheme repairs itself because integrated circuits without power are not sensitive to SEL. For the successful operation of this scheme, it is necessary to provide a health check of microcontrollers using their software and provide switching of the power supply of microcontroller in a case of failure. These problems we solve using the watchdog MAX6369 by Maxim; JK-trigger produced in the CIS and the Solid State Relay PRAC37S by Proton.
The microcontrollers have to provide impulses to MAX6369 every 5 seconds, generated by the software and hardware health check. In the case of failure, the watchdog gives an impulse to the T-trigger. T-trigger switches to the opposite state and starts power supply for the second microcontroller while powering off the first. T-triggers are also redundant. In the worst conditions, in the case of failure of the T-trigger, there is a possibility of both microcontrollers powered on at the same time. There are two communication lines between microcontrollers for software arbitration mechanism in such case.

The unit uses the SPI and 1-wire buses for communication with other devices inside the unit. Data storage is a FLASH memory AT45DB011D by Atmel with the SPI interface. There are two MEMS 3-Axis gyroscope and the 3-Axis accelerometer MPU-9250 by InvenSense which provide a measurement of a spinning precession.

The deployment of the sail is provided by two bipolar stepper motors, as in the project BMSTU-Sail. Stepper motors are the most suitable for usage in the space due to the lack of brushes and possibility of control of spinning angle without feedback. In each coil, we used the bipolar stepper motor AM1524 by Faulhaber. To provide self-retardation, we used the reduction drive with a gear ratio more than 40.
Motors are less capable of operation in the low-temperature  (from -20$\degree C$) that's why they may need heating. For this purpose, there are two resistors connected with the thermal sensors.

Control of the motor is performed with the use of the dual H-bridge. They has wide temperature range: from - 60$\degree$C to +125$\degree$C. We use four digital signals from microcontroller's output ports to control sail deploying or stowing.

The tolerance to one failure is provided by the redundancy only of a driver because the probability of a motor hardware failure is negligible. The second H-bridge chip is connected parallel to the primary chip and stays in the cold reserve.
All chips have independent power switches.

The detailed analysis and experiments have shown that to provide full redundancy it needs to include the separate relays in each of four circuits from K1128KT4BR output to the stepper motor coils.

The operational tests of the primary and reserve chips and switching between them have been carried out on the breadboard. The drive has shown good fault tolerance on the accidents such as the excess of the maximum current, break of several operating wires, break up from the wires to the motor.

Two photosensors AOT137A1 from Optron control the sail deployment process. These sensors check if the bobbin is rotating or not. Sensors are located opposite to the cogwheels connected to shafts and detect the presence or absence of the windows in gear wheels. If they do not provide any changes in data at the specific time, the primary driver of this motor is considered failed, and the reserve driver turns on. A pair of sensors is necessary for ensuring the tolerance to one failure of the photosensors.

There are JTAG, UART and USB interfaces for debugging and programming.

The PCB design has to satisfy these requirements: coupling nuts have to be isolated from electrical circuits; bobbins bodies metallization has to be connected with a metallization circuit on PCB; electrical components have to be glued to PCB with thermally conductive glue; the area near high-power electrical components has to be covered with thermally conductive glue. 

All electric schematic is developed in Dip Trace EDA/CAD software for creating schematic diagrams and printed circuit boards.

\section{Qualification}

To prove the design of solar sail unit we need an extensive on-ground testing: electrical and functional testing, tests of software, tests of deployment of sail, vibration testing, climatic, thermal vacuum, and other tests.
To carry out these test, we need the working mock-up that have to be similar to flight unit. 
The ground qualification will be conducted in the Bauman MSTU.    
The partial space qualification of two-blade solar sail technology will be carried out during BMSTU-Sail space experiment~\cite{bmstu_iac_2015}.

We also proposed a mission which will involve three CubeSats with the solar sail. Each nanosatellite will be equipped with a spectrophotometer to carry out the observation of the Sun. One of the satellites should always be on the illuminated side of the orbit to ensure the continuous Sun observation. Thus, taking into account that all three nanosatellites will be launched from the same rocket booster at the same time, we need a proposed solar sailing device to create and to maintain the constellation.
This project started in 2017 as a research collaboration between Bauman Moscow State Technical University and the P.N. Lebedev Physical Institute of Russian Academy of Science, Moscow, Russia, and currently is in the preliminary phase.

\section{Conclusion}

In this work, we showed the feasibility and technological details of a method of the deployment of nanosatellite constellations using solar sails.
Solar sails, especially heliogyro-like, are still unproven to be reliable for space operations, but the calculation results showed that it is possible to deploy and maintain the constellation of small CubeSat class satellites on the particular range of orbit altitudes.

The advantage of this deployment method is the fact that this approach is faster than passive phase separation, and may be cost-effective relative to traditional space propulsion systems.
The other advantage of the described device is a selected type of solar sail -- it is possible to roll back the solar sail blade onto the bobbin for future use. However, this possibility is still unproven to be achievable in space flight.
It is planned to deploy and roll back the similar sail during the BMSTU-Sail Space Experiment.
The disadvantage of using a heliogyro-like solar sail is the fact that for several payloads, e.g., Earth remote sensing payload, it may be impossible to operate since the satellite should rotate around some axis.

This unit can increase the effectiveness of CubeSat satellite constellations for real-time space weather, Earth observation, improving life quality of people all over the world.

\section*{Acknowledgment}

Authors would like to thank Department ``Aerospace Systems (SM-2)'' of Bauman MSTU for granting access to their computer cluster, department ``Technologies of rocket engineering (SM-12)'' of Bauman MSTU and A.N. Korolev from this department for help with polyamide 3D printing.

The research was performed at Bauman Moscow State Technical University with the financial support of the Ministry of Education and Science of the Russian Federation under the Federal Target Program "Research and development on priority directions of the scientific and technological complex of Russia for 2014-2020". Agreement \# 14.574.21.0146 (unique identifier RFMEFI57417X0146).
	
	\bibliographystyle{apa} 	
	\bibliography{mybibfile}
	
\end{document}